\documentclass[english,twocolumn,showkeys]{revtex4}
\usepackage{mathptmx}

\usepackage[T1]{fontenc}
\usepackage[latin9]{inputenc}
\usepackage{color}
\usepackage{textcomp}
\usepackage{amsmath}
\usepackage{amssymb}
\usepackage{graphicx}

\makeatletter

\usepackage{babel}

\makeatother

\usepackage{babel}

\newcommand{\Bremsstrahlung}{Brems\-strahl\-ung }

\begin{document}

\title{Linear response of a variational average atom in plasma : semi-classical
model}

\author{C. Caizergues\footnote{clement.caizergues@cea.fr}, T. Blenski}

\address{CEA Saclay, IRAMIS, ``Laboratoire Interactions Dynamique et Lasers'', Centre d'Etudes de Saclay, F-91191 Gif-sur-Yvette
Cedex, France.}

\author{R. Piron}

\address{CEA, DIF, F-91297 Arpajon, France.}

\begin{abstract}

The frequency-dependent linear response of a plasma is studied in the finite-temperature Thomas-Fermi approximation, with electron dynamics described using Bloch hydrodynamics. The variational framework of average-atoms in a plasma is used. Extinction cross-sections are calculated for several plasma conditions. Comparisons with a previously studied Thomas-Fermi Impurity in Jellium model are presented. An Ehrenfest-type sum rule, originally proposed in a full quantum approach is derived in the present formalism and checked numerically. This sum rule is used to define \Bremsstrahlung and collective contributions to the extinction cross-section. It is shown that none of these is negligible. Each can constitute the main contribution to the cross-section, depending on the frequency region and plasma conditions. This result obtained in the Thomas-Fermi-Bloch case stresses the importance of the self consistent approach to the linear response in general. Some of the methods used in this study can be extended to the linear response in the quantum case.

\end{abstract}
\maketitle

\section{Introduction}

Modeling of dense plasma optical properties is necessary in order to calculate opacity and conductivity, widely needed in inertial confinement fusion, and astrophysics. For this purpose, Average-Atom (AA) models constitute often a first approach thanks to their relative simplicity and low computing cost. We may use AA models to study problems relative to the treatment of the plasma environment including screening and other density effects. Among the main difficulties in these problems is the presence and the appropriate treatment of the "free" or delocalized electrons and of the non-central ions. AA models are good testbed for such studies. We mean here especially the thermodynamic coherence of the models of atoms in plasma. The main issue here is the formulation of models in a correct variational framework  taking into account the ionization degree i.e. the number of delocalized electrons per atom as a  variational variable. 

During recent years  a progress has been obtained in this subject. It resulted in the "Variational Average-Atom in Quantum Plasma" (VAAQP) approach which was used in  both the non-relativistic and relativistic AA  as well as in the superconfiguration in plasma model \cite{Blenski07b,Blenski07a,Piron09,Piron09c,Piron11,Piron2013,Blenski2013}. It was also shown that in the case of the quasi-classical description of electron density the VAAQP approach led to the classical Thomas-Fermi (TF) atom at finite temperature proposed in \cite{Feynman49}.  In all these cited references the  ion correlation was accounted for in a heuristic way, typical for all preceding AA models (and also in \cite{Rozsnyai72,Liberman79,Blenski1997,Pain2003,Pain2006}) in which ion correlation has been including in the form of the Heaviside function corresponding to the Wigner-Seitz cavity (see \cite{Liberman79} and the discussion in \cite{Blenski2013}). The inclusion of non-central ions into a general variational scheme still remains an open problem despite some interesting ideas suggested in \cite{Starrett12}.

The AA in plasma models can be used to study frequency-dependent electron properties. This can be done in the framework of the dynamic Linear Response (LR) theory of the AA variational equilibrium. Such studies may give information whether and at which plasma condition the collective electron dynamic phenomena may be important. This  may have practical implications since the absorption and resistivity can be calculated directly from the LR theory. An important problem is the LR in quantum models treating all electrons on the same footing. The LR frequency-dependent theory of the AA quantum model has been proposed in \cite{Blenski92} and in \cite{Blenski06}. The idea was to use the formalism of the cluster expansion first introduced in the case of polarizable suspensions \cite{Felderhof82}. The cluster expansion formalism allows one to solve the problem of localization of the AA response by the  subtraction of the homogeneous plasma response.

The frequency-dependent LR based on the cluster expansion in the case of the quasi-classical TF ion immersed in plasma has been investigated in details in a series of publications \cite{Felderhof95a,Felderhof95b,Felderhof95c,Ishikawaall1998,IshikawaThesis,IshikawaFelderhof1998}. The problem was the finite temperature extension of the classical paper \cite{Ballall1973}, that was motivated by the earlier results of \cite{Kirzhnits1966,Kirzhnits1975} on the collective oscillations in atoms. The electron dynamic in these studies was based on the Bloch hydrodynamics leading to the Thomas-Fermi-Bloch (TFB) approach.  As shown in \cite{Felderhof95a,Felderhof95b,Felderhof95c,Ishikawaall1998,IshikawaThesis,IshikawaFelderhof1998}, the cluster expansion automatically cancels some divergent terms in free electrons contribution appearing due to the fact that free electrons do not belong to atoms but to the plasma as a whole. In this way the cluster expansion approach leads formally to finite expression for the atomic response. However, the TFB LR studied in these publications can not be considered as the LR of the Thomas-Fermi AA but rather as the response of an impurity ion immersed in hot dense plasma \cite{Popovic74, Manninen77}. The important contribution of this series of papers on the TFB LR theory and its applications  was the understanding of the general LR formalism and the development of an original method allowing one to solve the first order TFB LR equations taking into account the asymptotic behavior of the first order quantities. This method is highly efficient and provides directly the induced atomic dipole. It gives access to the general aspects the frequency-dependent LR theory of atoms in plasma that are important for future possible quantum extensions of the LR theory. 

In the present paper we report the first calculation of the frequency-dependent LR of the finite temperature Thomas-Fermi AA of \cite{Feynman49} considered in the framework of the VAAQP approach. The fact that the VAAQP variational approach leads in an unique way to the classical Thomas-Fermi AA from \cite{Feynman49} was  understood in \cite{Blenski07b}. The method we use to solve the TFB set of equations is the same as the one proposed in \cite{Ishikawaall1998}. 

The AA TF equilibrium is presented in section~\ref{TF}. The cluster expansion and the TFB set of equations are derived in section~\ref{LR}. In section~\ref{sec:sum_rule} we prove that the Ehrenfest-Type Atom-in-Plasma Sum Rule (ETAPSR) which has been previously derived in the quantum AA case \cite{Blenski2013} remains valid in the TFB LR model. The numerical solution of the LR TFB set of equations allows us for the first time to check this sum rule and discuss its physical meaning. The numerical results for the extinction cross section are analyzed in section~\ref{sec:Photoabsorption in TFB}. The VAAQP formalism introduces the presence of the WS cavity, which is neutral in the Thomas-Fermi case. This has an impact on the values of the AA absorption cross-section especially close to the plasma frequency. The ETAPSR allows us also to distinguish two terms in the induced atom dipole and in the extinction cross-section. The first we call the "\Bremsstrahlung" like term since it leads to the \Bremsstrahlung term in the independent electron approach (see, for example, \cite{Johnson06}) and the   second the "collective" like term. The  contribution of these two terms to the extinction cross-section in function of plasma parameters and frequency range is discussed. The conclusions are given in section~\ref{Conclusion}.

\section{Equilibrium description: Thomas-Fermi atom}
\label{TF}

An approach to plasma modeling consists of treating locally electrons as an ideal Fermi gas. The electron density $n(\vec{r})$ then only depends on the local potential value $\phi(\vec{r})$. This is the usual Thomas-Fermi (TF) hypothesis. In the considered non-relativistic case:
\begin{equation}
n(\vec{r})=2\int\frac{d\vec{p}}{h^3}\frac{1}{\exp\left(\beta\left(\frac{p^2}{2m}-\mu_{0}-e\phi(\vec{r})\right)\right)+1}.\label{Fermi gas}
\end{equation}
Here, $\mu_0$ stands for the chemical potential, $\beta$ is the inverse of the temperature $1/k_{B}T$, $m$ is the electron mass, and $e$ is the electron absolute charge.
 
A first approach of this kind to atoms in condensed matter appeared in \cite{Slater35}. In this reference, the Wigner-Seitz (WS) polyhedron cell resulting from the periodic structure of metal at zero temperature was replaced by a sphere of radius $r_{WS}$ such that:
\begin{equation}
\frac{4}{3}\pi r_{WS}^{3}=\frac{1}{n_{i}}.\label{eq:rayon WS}
\end{equation}
where $n_i$ is the atom density. The application of a TF atom contained in the WS sphere to compressed matter and finite temperatures plasma was first proposed in \cite{Feynman49}. In both approaches, the equations for the self consistent electron density and electrostatic potential are as follows:
\begin{align}
&\nabla^{2}\phi(\vec{r})=4\pi en(\vec{r}),\label{eq:Poisson}\\
\phi(r\rightarrow0)=\frac{Ze}{r}&,\ \phi\left(r_{WS}\right)=0,\ \left.\frac{d\phi}{dr}\right|_{r=R_{WS}}=0,\label{eq:boundary}
\end{align}
where $n(\vec{r})$ is given by \eqref{Fermi gas}. The boundary conditions \eqref{eq:boundary} account for the central nuclear charge $Ze$ and the neutrality of the WS sphere. In what follows, we will call this model Thomas-Fermi Average-Atom (TFAA). In the TFAA model, the equilibrium is determined by three parameters: $n_i$, $Z$, $T$.

The TFAA has been widely used in the literature because of its simplicity, thermodynamic consistency, and scaling with respect to the atomic number $Z$. The results obtained using the TFAA model for a given element can be generalized to others thanks to the scaling law in $Z$ of the model. Equations can be made independent of $Z$, if written using the following four quantities:
\begin{equation}
TZ^{-4/3},nZ^{-2},\phi Z^{-4/3},rZ^{1/3}.\label{Scaling_EOS}
\end{equation}

At finite temperatures, the physical picture of the WS sphere has to be interpreted differently from the idea of \cite{Slater35} since in plasmas, there is no periodical structure. It stems from the TFAA model that a finite electron density $n_0^{(0)}=n(r_{WS})$ related to the chemical potential $\mu_0$ is obtained at the WS boundary. One may then consider as in \cite{Liberman79} that beyond the WS sphere is a jellium of electron density $n_0^{(0)}$. This implies the presence beyond the WS sphere of a homogeneous, neutralizing non-central ion background composed of ions of effective charge $Z^*=n_0^{(0)}/n_i$. These non-central ions then have a charge density: $\rho_{i}\left(r\right)=n_{0}^{(0)}e\Theta\left(r-r_{WS}\right)$, which corresponds to a WS cavity.

This picture of one ion in a jellium with a cavity was used in the framework of a cluster expansion (see \cite{Felderhof82}) in previous works leading to the Variational Average-Atom in Quantum Plasma (VAAQP) model (see \cite{Blenski07a, Blenski07b, Piron11}). It was proved that the TFAA model can also be considered as resulting from the VAAQP approach, if the electron free-energy is taken in the TF approximation. This interpretation is the starting point of the present study.

In \cite{Ishikawaall1998} another TF model at finite temperature is studied, corresponding to an impurity in a jellium of a given electron density $n_0^{(0)}$, without any WS cavity. The jellium ion charge density is then $\rho_{i}=en_{0}^{(0)}$. We will call this model Thomas-Fermi Impurity in Jellium (TFIJ). Instead of \eqref{eq:boundary} the boudary conditions become in the TFIJ case:
\begin{eqnarray}
\phi(r\rightarrow0)=\frac{Ze}{r}, & \phi(r\rightarrow\infty)=0.\label{eq:boundary_TFIJ}
\end{eqnarray}
The equations \eqref{Fermi gas}, \eqref{eq:Poisson}, and \eqref{eq:boundary_TFIJ} have to be solved in this case in the whole space. The solution is determined by the three parameters: $n_0^{(0)}$, $Z$, $T$. The plasma ion density $n_i$ is absent in the TFIJ model.

In the present work, we will address the LR of the two equilibrium model TFAA and TFIJ. We will pay special attention to the TFAA model since it is well-suited to the description of plasmas. Moreover it constitutes a first step towards the application of the LR theory to the full quantum version of the VAAQP model. However for comparison and validation of our calculations we will also consider the TFIJ model, since several numerical results were obtained in this case \cite{Ishikawaall1998,IshikawaThesis}.

\section{Linear response: Bloch theory for an average-atom}
\label{LR}

To study the dielectric response of the TFAA and TFIJ models we use a hydrodynamic approach to electrons. We first consider a system consisting of $N$ ions and $ZN$ electrons, where the $N$ ions are supposed to be point-like charges fixed at the positions $\left\{\vec{X}_j\right\}$. The evolution of the electron fluid density $n^{(N)}(\vec{r},t)$, flow velocity $\vec{v}^{(N)}(\vec{r},t)$ and total potential $\phi^{(N)}(\vec{r},t)$, including an external time-dependent perturbation, is governed by the equation of continuity and the equation of motion:
\begin{align}
&\frac{\partial n^{(N)}}{\partial t}+\vec{\nabla}\centerdot(n^{(N)}\vec{v}^{(N)})=0,\label{eq:continuity}\\
nm&\frac{d\vec{v}^{(N)}}{dt}=-\vec{\nabla} p^{(N)}+n^{(N)}e\vec{\nabla}\phi^{(N)}.\label{eq:motion}
\end{align}
In~\eqref{eq:continuity} and \eqref{eq:motion}, all quantities also depend on the ion positions $\left\{\vec{X}_j\right\}$, although we do not write these dependence explicitly. In~\eqref{eq:motion} the pressure $p^{(N)}(\vec{r},t,T)$ is assumed to be locally related to the electron density by a known Equation Of State (EOS) $p(n,T)$.

As in the previous section, we will consider the non-relativistic Fermi gas \eqref{Fermi gas} and its corresponding EOS. Together with \eqref{eq:continuity} and \eqref{eq:motion}, this leads to Bloch hydrodynamics \cite{Bloch1933}. Notice that one could use the relativistic Fermi gas or add a local-density exchange-correlation contribution without changing the formalism presented here. To complete the system of hydrodynamic equations the Poisson equation should be added. This will account for the external time-dependent density of charge $\rho_{ex}(\vec{r},t)$ corresponding to the perturbing field and the point-like ion charge density $\rho_i^{(N)}\left(\left\{\vec{X}_j\right\},\vec{r}\right)$:
\begin{equation}
\nabla^{2}\phi^{(N)}=4\pi n^{(N)}e-4\pi\rho_i^{(N)}-4\pi\rho_{ex}.\label{eq:Poisson External}
\end{equation}

In the absence of external field the equation of motion leads to
\begin{equation}
\vec{\nabla} p_{0}^{(N)}(\vec{r})-n_{0}^{(N)}(\vec{r})e\vec{\nabla}\phi_{0}^{(N)}(\vec{r})=0,\label{eq:no field}
\end{equation}
which solution is:
\begin{equation}
\mu_{0}^{(N)}(\vec{r})-e\phi_{0}^{(N)}(\vec{r})=const,
\label{equilibre_mu}
\end{equation}
where $\mu_{0}^{(N)}(\vec{r})$ is defined from the EOS, as:
\begin{equation}
\mu_{0}^{(N)}(\vec{r})=\int^{p(n_0^{(N)})}\frac{dp}{n},
\label{def_mu} 
\end{equation}
or, in general:
\begin{equation}
\mu(n,T)=\int^{p(n,T)}\frac{dp}{n}.
\label{eq:EOS_int} 
\end{equation}
In order to find the linear dielectric response we consider a weak perturbation of the equilibrium, expand Bloch equations \eqref{eq:continuity}, \eqref{eq:motion} and \eqref{eq:Poisson External} to first order and use equation \eqref{eq:no field} in order to simplify:
\begin{align}
&\frac{\partial n_{1}^{(N)}}{\partial t}+\vec{\nabla}\centerdot(n_{0}^{(N)}\vec{v}_{1}^{(N)})=0,\\
&m\frac{\partial S_{1}^{(N)}}{\partial t}=-\mu_{1}^{(N)}+e\phi_{1}^{(N)},\\
&\nabla^{2}\phi_{1}^{(N)}=4\pi en_{1}^{(N)}-4\pi\rho_{ex},\label{eq:Poiss1}
\end{align}
where we make the hypothesis that the first order velocity flow is non-rotational: $\vec{v}_{1}^{(N)}=\vec{\nabla} S_{1}^{(N)}$ and $\mu_{1}^{(N)}$ is the first order chemical potential:

\begin{equation}
\mu_{1}^{(N)}=\vartheta_{0}^{(N)}n_{1}^{(N)},
\end{equation}
with the definition:
\begin{equation}
\vartheta_{0}^{(N)}=\left.\frac{\partial\mu}{\partial n}\right|_{n_{0}^{(N)}}=\frac{1}{n_{0}^{(N)}}\left.\frac{\partial p}{\partial n}\right|_{n_{0}^{(N)}}.\label{theta}
\end{equation}

We consider an harmonic perturbation in the form of a homogeneous electric field $\vec{E}_{ex}(\vec{r},t)=\vec{E}_{ex}(t)$, i.e. in the dipole approximation. Hence external charge disappear in \eqref{eq:Poiss1}. 
Introducing the Fourier transforms $n_{\omega}^{(N)}(\vec{r})$, $\phi_{\omega}^{(N)}(\vec{r})$, $S_{\omega}^{(N)}(\vec{r})$, $\vec{E}_{ex,\omega}$ with the definitions as for instance: 
\begin{align}
n_{\omega}^{(N)}(\vec{r})=\int e^{i\omega t}n_{1}^{(N)}(\vec{r},t) dt,\label{eq:fourier_transform}
\end{align}
and replacing these quantities in the linearized Bloch equations one finds:
\begin{align}
-&i\omega n_{\omega}^{(N)}+\vec{\nabla}\centerdot(n_{0}^{(N)}\vec{\nabla} S_{\omega}^{(N)})=0,\label{eq:cv_densite}\\
-&i\omega mS_{\omega}^{(N)}+\vartheta_{0}^{(N)}n_{\omega}^{(N)}-e\phi_{\omega}^{(N)}=0,\label{eq:cv_impulsion}\\
&\nabla^{2}\phi_{\omega}^{(N)}=4\pi en_{\omega}^{(N)}\label{eq:poisson}.
\end{align}

The set of equations \eqref{eq:cv_densite}, \eqref{eq:cv_impulsion} and \eqref{eq:poisson} applies to an electron fluid with fixed ion positions. In principle this set of equations have to be solved for each possible set of ion positions. The observables should then be obtained taking the average over all set of ion positions with a probabiblity distribution $W\left(\left\{\vec{X}_j\right\}\right)$. Fortunately such an average can be constructed iteratively by the cluster expansion proposed in \cite{Felderhof95a} and \cite{Felderhof95b}. To zeroth order the electron-ion plasma is approximated by an infinite and homogeneous jellium at finite temperature. In the first order, one gets the notion of an average atom immersed in a jellium. The first-order LR can be constructed by subtracting the response of an infinite plasma from the response of such an atom. The thermodynamic limit $N\rightarrow\infty$, $V\rightarrow\infty$, at constant $N/V=n_i$ is taken for each order. We limit ourselves to the two first orders and denote by $^{(0)}$ and $^{(1)}$ their corresponding quantities, respectively. Within this framework, the set of equations \eqref{eq:cv_densite}, \eqref{eq:cv_impulsion} and \eqref{eq:poisson} become:
\begin{align}
&n_{\omega}(\vec{r})=n_{\omega}^{(1)}(\vec{r}),\\
&\phi_{\omega}(\vec{r})=-\vec{E}_{ex,\omega}\centerdot\vec{r}+\phi_{\omega}^{(1)}(\vec{r}),\label{eqs19}\\
&S_{\omega}(\vec{r})=-i\frac{e}{\omega m}\vec{E}_{ex,\omega}\centerdot\vec{r}+S_{\omega}^{(1)}(\vec{r}).\label{eqs20}
\end{align}
One can note that, due to the symmetry and particle number conservation, $n_\omega^{(0)}(\vec{r})=0$ in the dipole approximation.

After introduction of the variable $\sigma_{\omega}=-i\omega mS_{\omega}/e$, the set of equations \eqref{eq:cv_densite}, \eqref{eq:cv_impulsion} and \eqref{eq:poisson} finally becomes: 
\begin{align}
&-e\omega^2 n_{\omega}^{(1)}+\frac{e^2}{m}\vec{\nabla}\centerdot(n_{0}\vec{\nabla}\sigma_{\omega}^{(1)})=\frac{e^2}{m}\vec{E}_{ex,\omega}\centerdot\vec{\nabla}n_{0},\label{RHO}\\
&\sigma_{\omega}^{(1)}=-\frac{\vartheta_{0}}{e}n_{\omega}^{(1)}+\phi_{\omega}^{(1)},\label{SIGMA}\\
&\nabla^{2}\phi_{\omega}^{(1)}-4\pi en_{\omega}^{(1)}=0.\label{eq:poisson1}
\end{align}

To solve this system one should take care of the spherical symmetry of the first order equilibrium model. The electron density of this model is denoted by $n_0(r)$. We choose the $z$-axis in the direction of the perturbing electric field. Using spherical coordinates $\left(r,\theta,\phi\right)$ and the variables proposed in \cite{Ishikawaall1998}: 
\begin{align}
&\sigma_{\omega}^{(1)}(\vec{r})=\frac{G(r)}{r\sqrt{n_{0}}}cos(\theta),\label{eq:chgt var sigma}\\
&\phi_{\omega}^{(1)}(\vec{r})=\frac{H(r)}{r}cos(\theta),\label{chgt var phi}\\
&n_{\omega}^{(1)}(\vec{r})=\frac{K(r)}{r}cos(\theta),\label{eq:chgt var n1}
\end{align}
leads to the set of equations:
\begin{align}
&\frac{d^2G}{dr^2}+a_{GG}G+a_{GH}H=S,\label{eq:diff G}\\
&\frac{d^2H}{dr^2}+a_{HG}G+a_{HH}H=0,\label{eq:diff H}\\
&K=\frac{e}{\vartheta_{0}}\left(H-\frac{G}{\sqrt{n_{0}}}\right).\label{eq:G H K}
\end{align}
Denoting by primes the derivatives with respect to the radial coordinate $r$, the coefficients and source term appearing in \eqref{eq:diff G} and \eqref{eq:diff H} are:
\begin{align}
&a_{GG}=-\frac{2}{r^{2}}-\frac{n_{0}^{'}}{rn_{0}}+\frac{(n_{0}^{'})^{2}} {4n_{0}^{2}}-\frac{n_{0}^{''}}{2n_{0}}+\frac{m\omega^{2}}{n_{0}\vartheta_{0}},\\
&a_{GH}=-\frac{m\omega^{2}}{\sqrt{n_{0}}\vartheta_{0}},\\
&a_{HH}=-\frac{2}{r^{2}}-\frac{4\pi e^{2}}{\vartheta_{0}},\\
&a_{HG}=\frac{4\pi e^{2}}{\sqrt{n_{0}}\vartheta_{0}},\\
&S=\frac{rn_{0}^{'}}{\sqrt{n_{0}}}E_{ex,\omega},
\end{align}

The differential set of equations can be solved using the known behavior of the equilibrium quantities at the boundaries. For small radii $n_{0}(r)$ is directly related to the nuclear charge. For the non-relativistic Fermi gas relation one gets the following behaviors:
\begin{align}
&G(r)\underset{r\rightarrow0}{\longrightarrow}\frac{E_{ex,\omega}}{\pi\sqrt{3}}\left(\frac{2mZe^{2}}{\hbar^{2}}\right)^{3/4}r^{5/4}+C_{G}r^{1/2+\sqrt{33}/4},\label{eq:G en 0}\\
&H(r)\underset{r\rightarrow0}{\longrightarrow}C_{H}r^{2}.\label{eq:H en 0}
\end{align}
In \eqref{eq:G en 0} and \eqref{eq:H en 0}, two initially unknown coefficients $C_{G}$, $C_{H}$ appear, which should be determined from the behavior of the induced quantities far from the central ion charge, where $n_{0}$ tends to $n_0^{(0)}$. For these large values of $r$, the functions $G$, $H$ and $K$ fulfill:
\begin{align}
&\frac{d^2K}{dr^2}+\left(-\frac{2}{r^{2}}-\alpha^{2}\right)K=0,\label{eq:asymptotic diff K}\\
&\left[\frac{d^2}{dr^2}-\frac{2}{r^{2}}\right]\left(H-\frac{\omega_{p}^{2}}{\omega^{2}\sqrt{n_{0}^{(0)}}}G\right)=0.\label{eq:asympotic diff H-G}
\end{align}
Here $\omega_{p}$ is the plasma frequency corresponding to the density $n_0^{(0)}$, namely:
$\omega_{p}=\sqrt{{4\pi n_0^{(0)}e^2}/{m}}$, and  $\alpha$ is such that:
\begin{equation}
\alpha^{2}=\frac{4\pi e^{2}}{\vartheta_{0}^{(0)}}\left(1-\frac{\omega^{2}}{\omega_{p}^{2}}\right)\label{alpha-1}
\end{equation}
where 
\begin{equation}
\vartheta_{0}^{(0)}=\frac{h^{3}\sqrt{\beta}}{4\pi\left(2m\right)^{3/2}F_{1/2}^{'}\left(\beta\mu_{0}\right)},
\end{equation}
with $F_{1/2}$ being the Fermi-Dirac integral of order $1/2$ (see, for instance, \cite{Ishikawaall1998}).

From now on, we are only interested in the frequencies $\omega>\omega_p$, since only in these cases the dipole approximation can be justified. From equations \eqref{eq:asymptotic diff K}, \eqref{eq:asympotic diff H-G} and \eqref{eq:G H K} it follows that the functions $K$ and $H$ behave for large $r$ as:
\begin{align}
&K(r)\underset{r\rightarrow\infty}{\longrightarrow}P_{K}ri_{1}(\alpha r)+Q_{K}rk_{1}(\alpha r),\label{eq:asymptotique K}\\
&H(r)\underset{r\rightarrow\infty}{\longrightarrow}\frac{4\pi e}{\alpha^{2}}K(r)+P_{H}r^{2}+\frac{Q_{H}}{r}.\label{eq:asympt H}
\end{align}
$i_{1}(z)$ and $k_{1}(z)$ are the first-order modified spherical Bessel functions of the first and third kinds, respectively. Following the definition of the Fourier transform \eqref{eq:fourier_transform}, the causality principle leads to $\omega$ having an infinitesimal positive imaginary part. One can retain the negative imaginary root for $\alpha$. Then, among the four coefficients in \eqref{eq:asymptotique K}, \eqref{eq:asympt H}, one shall be zero: $P_{K}$. $P_{H}$ shall also be zero because it leads to a diverging induced potential.

The correct solution can be found using the method proposed in \cite{Ishikawaall1998} where an approach is used to find the constants $C_{G}$, $C_{H}$ in \eqref{eq:G en 0} and \eqref{eq:H en 0} such that $P_{K}=P_{H}=0$ in the asymptotic solution far from the central ion charge. In order to do this one chooses three sets  of coefficients $C_{G,i},C_{H,i}$ and integrates outward the equations \eqref{eq:diff G} \eqref{eq:diff H} for each set. Far from the center in the asymptotic region the $P_{K,i}, P_{H,i} $ coefficients are determined for each set and the solution is built as a linear combination with the coefficients $a_i$, requiring:
\begin{align}
&a_{1}+a_{2}+a_{3}=1,\\
&a_{1}P_{K,1}+a_{2}P_{K,2}+a_{3}P_{K,3}=0,\\
&a_{1}P_{H,1}+a_{2}P_{H,2}+a_{3}P_{H,3}=0.
\end{align}
All unknown coefficients, including $Q_H$, which stand in the equations \eqref{eq:asymptotique K} and \eqref{eq:asympt H} are obtained taking the same combination using the coefficients $a_i$. The induced atomic dipole defined as
\begin{equation}
p_{\omega}^{(1)}=-e\int d\vec{r}zn_{\omega}^{(1)},\label{DIP}
\end{equation}
is obtained directly from the above procedure since one has
$p_{\omega}^{(1)}=Q_{H}$.

The photon energy extinction cross-section is then given by the formula \cite{LandauElectrodynamicsCont},
\cite{Felderhof95a}:
\begin{equation}
\sigma_{ext}^{(1)}(\omega)=\frac{4\pi}{c}\sqrt{\omega^{2}-\omega_{p}^{2}}\,\text{Im}\left(p_{\omega}^{(1)}/E_{ex,\omega}\right).\label{extinction}
\end{equation}
For the physical situations considered in dense plasma physics the scattering
cross section is usually negligible compared to the photoabsorption cross section (see, for instance, \cite{Ishikawaall1998}) and $\sigma_{ext}^{(1)}\approx\sigma_{abs}^{(1)}$.

\section{Sum rule}
\label{sec:sum_rule}
From the LR theory of section \ref{LR} one can derive a new sum rule relating the atomic dipole $p_{\omega}^{(1)}$ to the gradients of the equilibrium density and potential. From the definition of the atomic dipole \eqref{DIP}, using \eqref{RHO} one obtains:
\begin{equation}
p_{\omega}^{(1)}=\frac{e^2}{m\omega^2}\left[\int d\vec{r}zE_{ex,\omega}\frac{\partial n_{0}}{\partial z}-\int d\vec{r} z\vec{\nabla}\centerdot(n_{0}\vec{\nabla}\sigma_{\omega}^{(1)})\right].\label{SUM_1}
\end{equation}
We integrate \eqref{SUM_1} two times by part keeping the non-vanishing surface terms, which leads successively to:
\begin{align}
p_{\omega}^{(1)} = & \frac{e^{2}}{m\omega^{2}}\left[-\int d\vec{r}\phi_{ex,\omega}\frac{\partial n_{0}}{\partial z}-\int\vec{dS}\centerdot\left(zn_{0}\vec{\nabla}\sigma_{\omega}^{(1)}\right)\right.\nonumber\\
&\left.+\int d\vec{r}n_{0}\vec{e_{z}}\centerdot\vec{\nabla}\sigma_{\omega}^{(1)}\right]\nonumber \\
= & \frac{e^{2}}{m\omega^{2}}\left[-\int d\vec{r}\phi_{ex,\omega}\frac{\partial n_{0}}{\partial z}-\int\vec{dS}\centerdot\left(zn_{0}\vec{\nabla}\sigma_{\omega}^{(1)}\right)\right.\nonumber\\
&\left.+\int\vec{dS}\centerdot\left(\sigma_{\omega}^{(1)}n_{0}\vec{e_{z}}\right)-\int d\vec{r}\frac{\partial n_{0}}{\partial z}\sigma_{\omega}^{(1)}\right].\label{SUM_2}
\end{align}
The integral in \eqref{DIP} is over the whole space, so surface integrals in \eqref{SUM_2} must be evaluated in the limit of a surface infinitely far from the central ion. From \eqref{eq:chgt var sigma}, \eqref{eq:G H K},  \eqref{eq:asymptotique K} and \eqref{eq:asympt H} we know that far from the ion center $\sigma_{\omega}^{(1)}$ takes the following asymptotic form:
\begin{equation}
\sigma_{\omega}^{(1)}(\vec{r})\underset{r\rightarrow\infty}{\longrightarrow}\left(\frac{p_{\omega}^{(1)}}{r^2}+C_{\sigma} k_{1}(\alpha r)\right)cos(\theta).\label{SUM_3}
\end{equation}
where $C_{\sigma}$ is a constant.

In \eqref{SUM_3} the term proportional to $k_{1}(\alpha r)$ dominates the numerical solution at large radial values. Nevertheless the causality principle leads to the presence of an infinitesimal positive imaginary part in $\omega$, and to an infinitesimal positive real part in $\alpha$ (see \eqref{alpha-1}) so this term does not contribute to the surface integrals at $r\rightarrow\infty$. Using the substitutions:
\begin{align}
\sigma_{\omega}^{(1)}(\vec{r}) &\underset{r\rightarrow\infty}{\longrightarrow} \frac{p_{\omega}^{(1)}}{r^2}cos(\theta),\\
\vec{e_{r}}\centerdot\vec{\nabla}\sigma_{\omega}^{(1)}(\vec{r})&\underset{r\rightarrow\infty}{\longrightarrow}\frac{-2p_{\omega}^{(1)}}{r^3}cos(\theta),
\end{align}
the two surface terms give:
\begin{equation}
\frac{e^2}{m}\left(-\int_{\Sigma(r)}\vec{dS}\centerdot\left(zn_{0}\vec{\nabla}\sigma_{\omega}^{(1)}\right)+\int_{\Sigma(r)}\vec{dS}
\centerdot\left(\sigma_{\omega}^{(1)}n_{0}\vec{e_{z}}\right)\right)
\underset{r\rightarrow\infty}{\longrightarrow}\omega_{p}^{2}p_{\omega}^{(1)},
\label{eq:Surface}
\end{equation}
where $\Sigma(r)$ is the sphere of radius $r$. From the local relation between pressure, density and potential (see \eqref{equilibre_mu} and \eqref{theta}) we can further obtain,
\begin{equation}
\frac{\partial n_{0}}{\partial z}\vartheta_{0}=\frac{\partial n_{0}}{\partial z}\left.\frac{\partial\mu}{\partial n}\right|_{n_{0}}=\frac{\partial\mu}{\partial z}=e\frac{\partial\phi_{0}}{\partial z},
\end{equation}
which together with \eqref{SIGMA} gives the sum rule:
\begin{equation}
p_{\omega}^{(1)}=\frac{e^2}{m\left(\omega^2-\omega_{p}^{2}\right)}\left[\int d\vec{r}n_{\omega}^{(1)}\frac{\partial\phi_{0}}{\partial z}-\int d\vec{r}\frac{\partial n_{0}}{\partial z}\phi_{\omega}\right].\label{SRULE}
\end{equation}

First, it is important to note that this sum rule provides a way to calculate the atomic dipole $p_{\omega}^{(1)}$. Indeed, the atomic dipole can not be directly evaluated by integration of the induced electron density because of the $k_{1}(\alpha r)$ like behavior of the electron density for large $r$. For that reason the integral in \eqref{DIP} is only conditionally convergent for $\omega>\omega_{p}$. This means that this integral only exists if one takes into account the causality principle introducing an infinitesimal positive imaginary part in the field frequency, as said on the occasion of the calculation of the surface terms \eqref{eq:Surface}. As presented  in Section \ref{LR} the method of solution of the set of the Bloch equations \eqref{RHO}, \eqref{SIGMA} and \eqref{eq:poisson1}, proposed in \cite{Ishikawaall1998},  leads directly to the determination of the induced dipole. However this is done without the explicit use of the definition \eqref{DIP}, i.e. without any integration of the induced electron density. The question appears then of an independent evaluation  of the induced dipole in order to check the convergence of the method.

In this respect the sum rule \eqref{SRULE} provides a way to calculate the induced dipole $p_{\omega}^{(1)}$. Let us notice that both integrals standing in \eqref{SRULE} are absolutely convergent, due to the gradients of the screened equilibrium electron density and potential which vanish far from the ion center. The causality principle has been used in the derivation of \eqref{SRULE} where evaluation of the surface terms was possible thanks to the infinitesimal positive imaginary part of $\omega$. Nevertheless, since the integrals standing in \eqref{SRULE} are absolutely convergent they can be evaluated for real $\omega$.

In the case of the semi-classical model based on the Bloch hydrodynamics we derive here the sum rule \eqref{SRULE}  for the first time. This sum rule was originally derived in an average-atom in plasma model using a full quantum  formalism \cite{Blenski06}. The fact that one recovers this sum rule within the semi-classical formalism is important for two reasons. First, it suggests a possible universal character of the sum rule in the plasma LR theory. Its validity in both semi-classical and quantum approaches stresses the relation between these two frameworks. The second reason is related to the relative simplicity of the semi-classical case. This make possible to study the underlying physics and mathematics related to the LR formalism in the case where the relation between the induced density and potential are local. Some methods tested on the semi-classical model may be extended to the self-consistent non-local response that appears in the quantum approaches.

In  \cite{Blenski06} it was shown that  from the sum rule \eqref{SRULE} one may obtain a special case of the Ehrenfest theorem (see for example \cite{Bethe_Salpeter}) if the induced potential is neglected :
\begin{equation}
\left\langle \psi_{0j}\left|z\right|\psi_{0i}\right\rangle =-\frac{1}{m\omega^{2}}\left\langle \psi_{0j}\left|e\frac{\partial\phi_{0}}{\partial z}\right|\psi_{0i}\right\rangle.\label{eq:ehrensfest}
\end{equation}
where $\psi_{0j}$,$\psi_{0i}$ are one-electron equilibrium wave functions and $\phi_{0}$ is the equilibrium atomic potential. The electron response is that of the independent particles constructed using the whole set of states $\left\{\psi_{0i}\right\}$. The equation \eqref{eq:ehrensfest} is the well known equivalence relation between the dipole and  acceleration matrix elements that one uses, for instance, in the calculation of the inverse \Bremsstrahlung cross-section. In fact, when both wave functions in \eqref{eq:ehrensfest} belong to the continuum spectrum the integral on the LHS is conditionally convergent whereas the RHS is absolutely convergent which is analogous to our preceding remarks on the sum rule \eqref{SRULE}.

Hence, in the quantum approach, the sum rule appears as a generalization of the Ehrenfest theorem applied to an AA in the LR approximation and will be referred as Ehrenfest-Type Atom-In-Plasmas Sum Rule (ETAPSR) (see \cite{Blenski2013}). In the present semi-classical Bloch case the induced density and potential are intrinsically related, so neglecting $\phi_{\omega}^{(1)}$ without also neglecting $n_{\omega}^{(1)}$ seems impossible. It does not seem relevant to construct an independent particle response in the case of the Bloch hydrodynamics. Thus in the semi-classical case the identification of a form of the Ehrenfest theorem similar to \eqref{eq:ehrensfest} does not appear obvious.

The first term appearing in the RHS of the sum rule involves the derivative of the equilibrium potential  so it has the form corresponding to the electron \Bremsstrahlung process. The second term involves the product between the derivative of the equilibrium density and the total oscillating field. It contains in principle other contributions including collective plasma effects. However the clear identification between \Bremsstrahlung and collective effects can only be performed in a more rigorous quantum approach involving electron wave functions. For the purpose of simplicity in the rest of the paper we will refer to the first term in the RHS of \eqref{SRULE} as the \Bremsstrahlung-like term and to the second as the collective-like term. In particular, the result highlighted in section \ref{sec:Photoabsorption in TFB} that the collective effects can dominate \Bremsstrahlung contributions for some values of $\omega$ needs to be confirmed in a more elaborate quantum approach.

As discussed above  in the semi-classical case using the method proposed in \cite{Ishikawaall1998} one obtains directly (see section \ref{LR}) the induced dipole $p_{\omega}^{(1)}$. Hence, in the semi-classical case one can numerically test the sum rule comparing the RHS of \eqref{SRULE} to the dipole value stemming from that method. We checked the sum rule \eqref{SRULE} in our numerical calculations of the LR of the two atom-in-plasma models presented in section \ref{TF} : TFAA and TFIJ atoms. 

Let us recall that in the case of the TFAA the gradients of the equilibrium quantities $\phi_{0}$ and $n_{0}$ are zero outside of the WS cavity, thus integrals appearing in equation \eqref{SRULE} are restricted to the WS sphere. It follows that in the TFAA LR the asymptotic behaviors \eqref{eq:asymptotique K}, \eqref{eq:asympt H} of the induced density and potential are exactly obtained outside of the WS sphere. So that the dipole coefficient in the induced potential can be precisely estimated just beyond the WS sphere. We found that in the TFAA case the sum rule \eqref{SRULE} is numerically very well verified. The relative error  between the RHS of \eqref{SRULE} and the dipole value stemming from the iterative procedure can be made less than $10^{-8}$. 

In the TFIJ case the gradients of the equilibrium potential $\phi_{0}$ and density $n_{0}$ vanish exponentially for large $r$. The characteristic length of this spatial screening is of the order of $1/{q_{T}}$ where $q_{T}^{2}={4{\pi}{e^{2}}}/{\theta_{0}^{(0)}} $. In the low temperature limit $q_{T}$ tends to ${q_{0}}$, the inverse of the Thomas-Fermi radius $r_{TF}=1/q_{0}=(e^{2}E_{F}/(6\pi n_{0}^{(0)}))^{1/2}$ with ${E_{F}}$ being the Fermi energy. Thus in the case of TFIJ the sphere of integration of the RHS side of \eqref{SRULE} should have a radius much larger than ${1}/{q_{T}}$ in order to get the asymptotic form of the induced quantities. For  that reason the integration region in the TFIJ case should be usually much larger than that of the TFAA case. The contribution of the asymptotic region to the RHS of \eqref{SRULE} is zero only in the TFAA case. In the TFIJ case the gradients of the equilibrium quantities may not be negligible even at distances of several $ r_{TF}$ from the atom center. Inspection of \eqref{alpha-1} shows that ${\alpha}$ is larger than $ {q_{T}}$ for $\omega$ much higher than $\omega_{p}$ and smaller than ${q_{T}}$ for $\omega$ close to $\omega_{p}$. Thus for high $\omega$ induced quantities oscillate  with $r$ and one may expect that the contribution from the asymptotic region to the integrals  in this case is generally small with respect to the case of  $\omega$ close to $\omega_{p}$. For that reason in the latter case the integrals standing on the RHS in the sum rule \eqref{SRULE} may be subject to errors  even when a relatively large integration sphere is taken. In general the numerical precision we obtained in checking the sum rule was poorer in the TFIJ case, relative error being usually of the order of $10^{-4}$.

\begin{figure}
\begin{center}
\includegraphics[width=8.5cm]{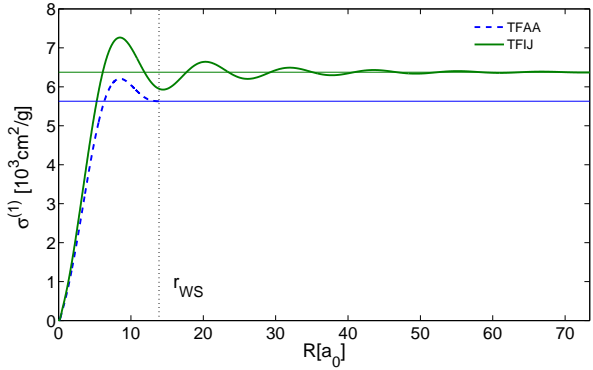}
\end{center}
\caption{(color online) Opacities  calculated from the RHS of the sum rule \eqref{SRULE} in function of the radius $R$ of the numerical region used for the integrals. Calculations are done in the case of TFAA and TFIJ for an aluminum plasma at $T=100eV$, $n_{0}^{(0)}=5.87 \times10^{21}\:cm^{-3}$ and $\omega=10\omega_{p}$. In the TFAA case this value of $n_{0}^{(0)}$ corresponds to the ion mass density of $0.027 gcm^{-3} $ i.e. one hundredth of the solid density. Constant curves correspond to the two respective cross sections obtained using the induced dipole provided by the procedure of section \ref{LR}. In the case of TFAA the sum rule leads to the same results as this procedure when the integral radius becomes equal to the WS radius whereas the TFIJ case requires much larger numerical radius of integration.\label{fig:Sum Rule}}
\end{figure}

On the figure \ref{fig:Sum Rule} we display the extinction cross section obtained using the dipole calculated from the sum rule \eqref{SRULE}. The calculations in both the TFAA and the TFJI cases concern an aluminum plasma at $T=100 eV$ temperature, with plasma (asymptotic) electron density $n_{0}^{(0)}=5.87\times 10^{21}\,cm^{-3}$ and ${\omega}=10\,\omega_{p}$. The radius of numerical sphere  used in the method of integration of the Bloch equations \eqref{RHO}, \eqref{SIGMA} and \eqref{eq:poisson1} was $ 10\,r_{TF}$. Both cross-sections on figure \ref{fig:Sum Rule} are displayed in function of the radius of the sphere used in the numerical integration of the RHS of the \eqref{SRULE}. In the two cases we also display by lines the cross-sections calculated using the induced dipole stemming from the  procedure described in section \ref{LR}. The necessity to use a greater radius of integration to calculate from the sum rule \eqref{SRULE} the atomic dipole in the TFIJ case appears clearly on the figure \ref{fig:Sum Rule}.

The numerical calculations presented in this article are all carried out in the case of Bloch hydrodynamics, i.e. for the EOS obtained from the non-relativistic Fermi gas relation. However this particular EOS is not required to establish the Ehrenfest-type sum rule \eqref{SRULE} which is more general. The derivation of \eqref{SRULE} presented above may be performed in the case of the classical hydrodynamics in the linear approximation using other local relation between pressure, density and potential. The sum rule remains valid for instance also in the case of small macroscopic polarizable particles immersed in a jellium (see \cite{LandauElectrodynamicsCont}). The optical wavelength however has to be larger that the particle radius in order to justify the dipole approximation. Besides in the case of atoms in plasma using the Fermi EOS one can easily incorporate local exchange-correlation effects without changing the formalism and the derivation of the sum rule \eqref{SRULE}. It is also interesting to notice that the sum rule \eqref{SRULE} has the same form both in our semi-classical  and in the quantum formalism \cite{Blenski06}, while these approaches are quite different. The semi-classical TFAA model has a local dependence between pressure, density and potential and the EOS clearly enters in the formalism. In the same time the relation between the induced density and potential are non-local in the quantum model and it is not necessary to make use of the EOS to establish the sum rule.

\section{Photo-absorption in Thomas-Fermi-Bloch case}
\label{sec:Photoabsorption in TFB}

In this section we present more comparisons of  photon extinction cross-sections between the two models : TFAA and TFIJ. For the purpose of simplicity we restrict our calculations to aluminum plasmas however one can extend  our conclusions  to plasmas of other elements using the scaling properties. Inspection of the LR set of equations (see \eqref{RHO}, \eqref{SIGMA} and \eqref{eq:poisson1}) and recalling the equilibrium TF scaling \eqref{Scaling_EOS}) gives the following scaling in the case of the Thomas-Fermi Bloch LR: 
\begin{equation}
\omega Z^{-1},E_{ex,\omega}Z^{-5/3},n_{\omega}^{(1)}Z^{-2},\phi_{\omega}^{(1)}Z^{-4/3},\sigma_{\omega}^{(1)}Z^{-4/3}\label{Scaling_LR}
\end{equation}
In the case of the TFAA model $ n_{i}$ respects the scaling  ${n_{i}}Z^{-1}$ that stems from its relation to the WS radius \eqref{eq:rayon WS}.

\begin{figure}
\centering{\includegraphics[width=8.5cm]{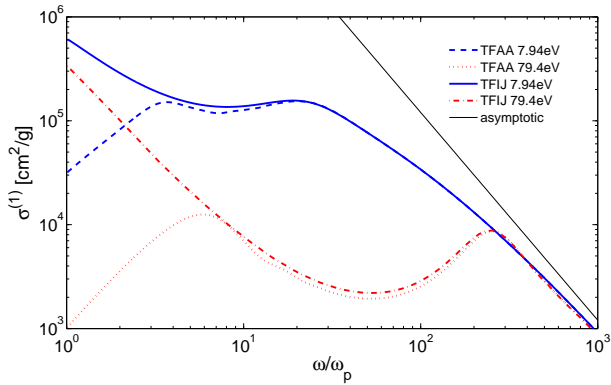}}
\caption{(color online) Comparison of opacities of aluminum plasma in function of frequency
for two temperatures $T_{1}=7.94\,eV$ and $T_{2}=10\,T_{1}$ and the two models: TFAA and TFIJ. The homogeneous plasma electron density is identical in the two cases $n_{0}^{(0)}=2.11 \times10^{21}\:cm^{-3}$. The two chosen plasma cases correspond to calculations performed in reference \cite{Ishikawaall1998} (fig. 10-11) where normalized Thomas-Fermi unit had been used. In the TFIJ cases we recover the same results. TFAA and TFIJ curves tend to the same results at high frequency but have quite different behavior near the plasma frequency.\label{fig:SectionComparison}}
\end{figure}

A first comparison is displayed on figure \ref{fig:SectionComparison} where opacities are plotted as functions of frequency (in $\omega_{p}$ units) for two  temperatures. Plasma parameters were chosen to be identical to those considered in \cite{Ishikawaall1998} where the TFIJ model was studied. Indeed in the TFIJ case we recover results from \cite{Ishikawaall1998}. 

As discussed in  \cite{IshikawaThesis} and \cite{IshikawaFelderhof1998} (see also \cite{Ballall1973}) at high frequency the cross-section behavior is connected to the spatial divergence of the Thomas-Fermi electron density induced by the point-like nuclear charge at the origin. This behavior is independent of the values of the electron density far from the atom center and consequently is independent of the chosen ion correlation function. Both models give then similar results and cross-sections follow the asymptotic behavior for large $\omega$ derived in~\cite{IshikawaFelderhof1998}: 
\begin{equation}
\sigma_{ext}^{(1)}(\omega)\propto 1/\omega^2.
\end{equation}

For frequencies close to the plasma frequency the cross-section values obtained from the TFAA and from the TFIJ models significantly differs. Contrary to the TFIJ case  the TFAA cross-section is an increasing function of frequency  near $ \omega_{p}$. In the TFIJ case, the total central charge $Z$ is screened by the equilibrium plasma electron density whereas in the TFAA case, this density screens only the charge $Z-Z^{*}$. As said before, for high frequencies the electron fluid close to the atomic center determines the absorption cross-section. On the contrary the electron density of the outer atomic region is mostly involved in the absorption process at lower frequencies. In this region the electron density is higher in the TFIJ model than in  the TFAA model. This is the reason for the higher absorption cross-sections in the TFIJ case.

On figure \ref{fig:SectionComparison} we compare opacities for two temperatures, at fixed $n_{0}^{(0)}$. In the TFAA model, the larger temperature $T_{2}$ corresponds to a lower ion density $n_{i}$ (see the $Z^{*}$ definition, section \ref{TF}). This implies larger WS radius and smaller gradients of the equilibrium density far from the atom center. According to the sum rule \eqref{SRULE} that leads to smaller absorption cross-sections close to plasma frequency. At high temperatures gradients are also smaller in this region in the TFIJ case. Indeed in both cases the opacities are lower at the highest temperature.

\begin{figure}[h!]
\centering{
\includegraphics[width=8.5cm]{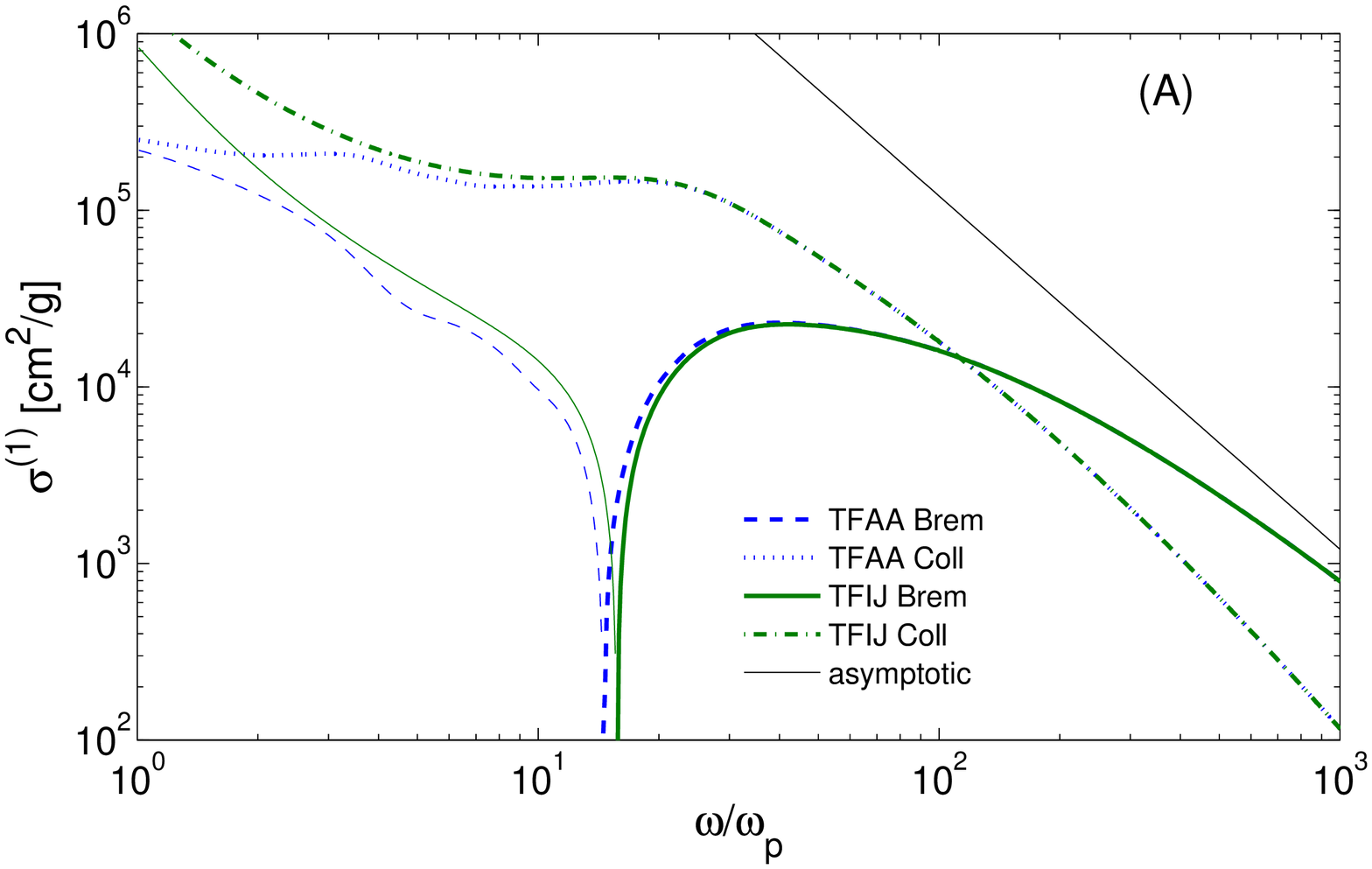}
\includegraphics[width=8.5cm]{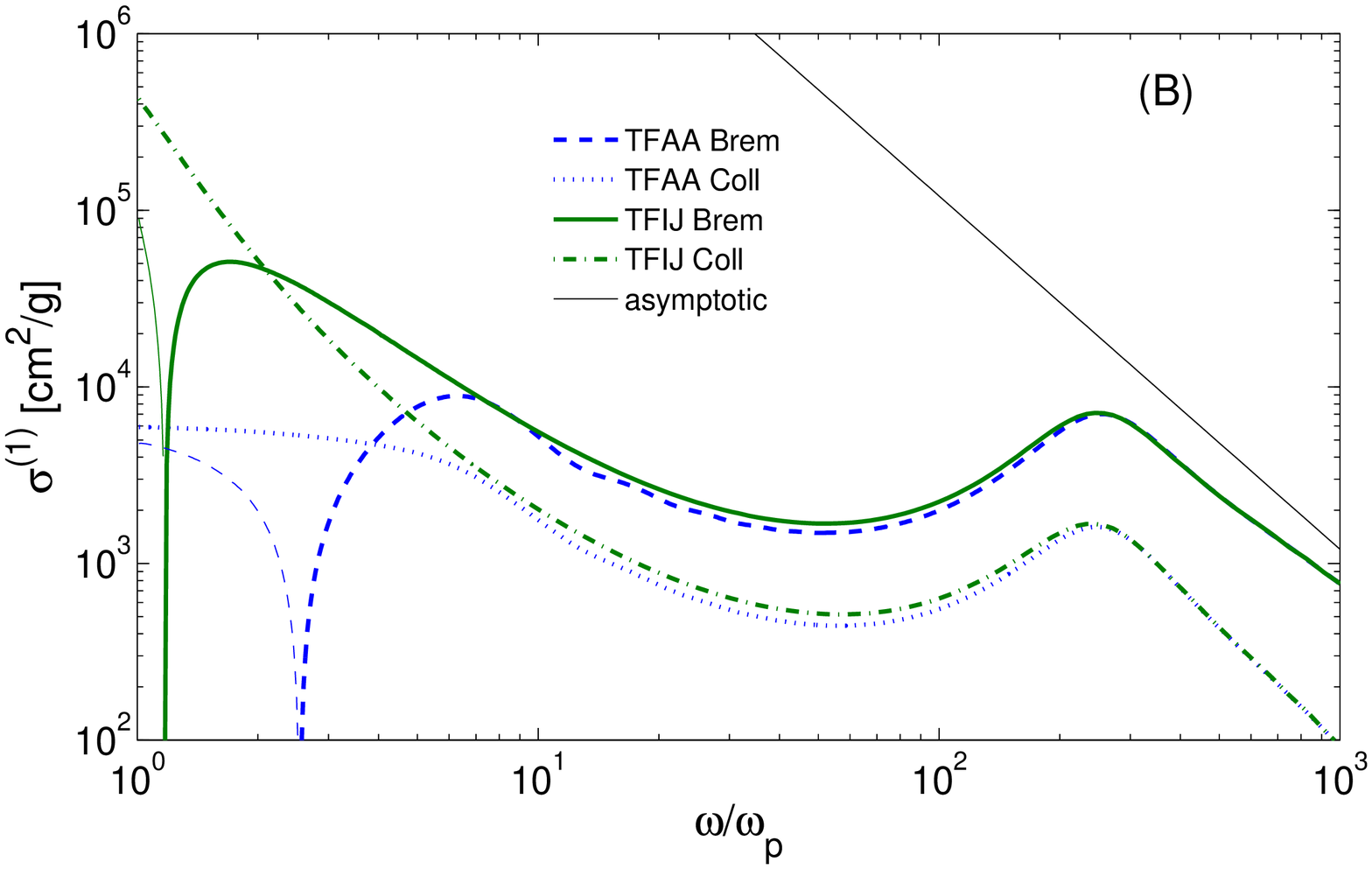}
}
\caption{(color online) Comparison between \Bremsstrahlung and collective term contributions to opacities  in the TFAA and TFIJ models. The plasmas conditions are the same as in figure \ref{fig:SectionComparison} : Figure A and B correspond to the temperature $T_{1}$ and $T_{2}$, respectively. At frequencies close to $\omega_{p}$ the \Bremsstrahlung terms become negative and their absolute values are displayed. At high frequency total opacities are dominated by the \Bremsstrahlung terms. The collective terms  seem to decay as $1/\omega^{3}$. 
\label{fig:ComparisonBremColl}}
\end{figure}

In the case of the TFIJ model, it was proved that the first order extinction cross section is finite at $\omega=\omega_p$ (see \cite{Felderhofall21995}). Looking at the sum rule \eqref{SRULE}, one finds immediately that near the plasma frequency, the sum of the integrals on the RHS should cancel in order to suppress the singularity of the cold plasma dielectric function. Indeed, there is no such singularity on the LHS of the sum rule. As already mentioned in section~\ref{LR}, the cluster expansion leads to a first-order response free of the plasmon divergences. Such divergences only appear in the zeroth order (i.e. homogeneous plasma) LR and are described in the dipole approximation by the cold plasma dielectric function. 

Using the sum rule \eqref{SRULE} and the extinction cross-section formula \eqref{extinction} one can try to identify the "\Bremsstrahlung" and the "collective"  parts entering in the atomic dipole and define the corresponding cross-sections. Such obtained \Bremsstrahlung and collective-effects opacities are presented on figure \ref{fig:ComparisonBremColl} in function of frequency in $\omega_{p}$ units. The plots correspond to the two temperatures of figure \ref{fig:SectionComparison}. As expected from standard theories, at high frequencies the \Bremsstrahlung cross-sections are larger than the collective ones. The \Bremsstrahlung term leads to the $1/\omega^{2}$ behavior of total extinction cross-section while collective effects seem to follow a $1/\omega^{3}$ decay. 

At low frequencies the relative contribution of the collective term increases and, in all of the considered cases, dominates over the \Bremsstrahlung term, starting from a frequency we call $\omega_B$. In all cases, at frequencies close to plasma frequency the \Bremsstrahlung term becomes negative, at a frequency we call $\omega_A$. In such situations this term may be interpreted as dominated by ``stimulated emission''. We recall however that we have defined the \Bremsstrahlung term in a arbitrary way in order to recover an expression similar to the usual \Bremsstrahlung term encountered in independent particle, i.e. non-self-consistent, approaches. Independently of this physical interpretation of the \Bremsstrahlung term, at low frequencies the two terms on the RHS of \eqref{SRULE} are of different sign with the same order of magnitude, which stresses possible limitations of the independent particle approximation in this frequency region.

\begin{figure}[h!]
\centering{
\includegraphics[width=8.5cm]{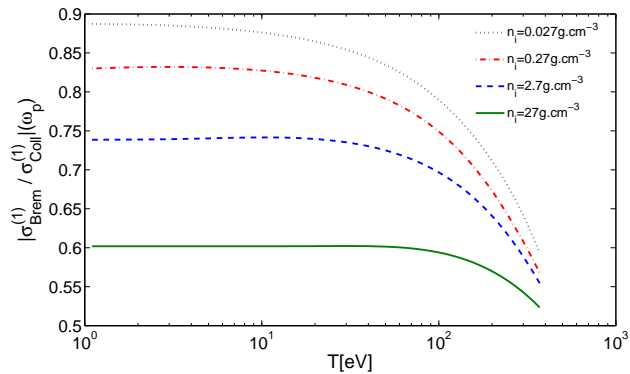}
}
\caption{(color online) The ratio between the \Bremsstrahlung to collective contribution to the opacity at $\omega=\omega_{p}$ in function of temperature for four ion mass densities. At low temperatures the ratio becomes constant and decreases with ion
mass density.
\label{fig:ratiowp}}
\end{figure}

To go further in our comparison between \Bremsstrahlung and collective effects we study their respective contributions to  the opacity for a wide range of plasma physical parameters. Here, the calculations have only been performed for the TFAA model, since it is the semi-classical version of the quantum approach to atoms in plasmas VAAQP (see \cite{Blenski07b, Blenski06, Blenski2013}). 

On figure \ref{fig:ratiowp} we display the absolute ratio between the \Bremsstrahlung and collective terms at the plasma frequency, as a function of temperature, for  four different ion densities. This ratio remains of the order of unity at temperatures between  $1eV$ and approximately $400eV$ for the ion densities considered. It however decreases at higher temperatures.

\begin{figure}[h!]
\centering{
\includegraphics[width=8.5cm]{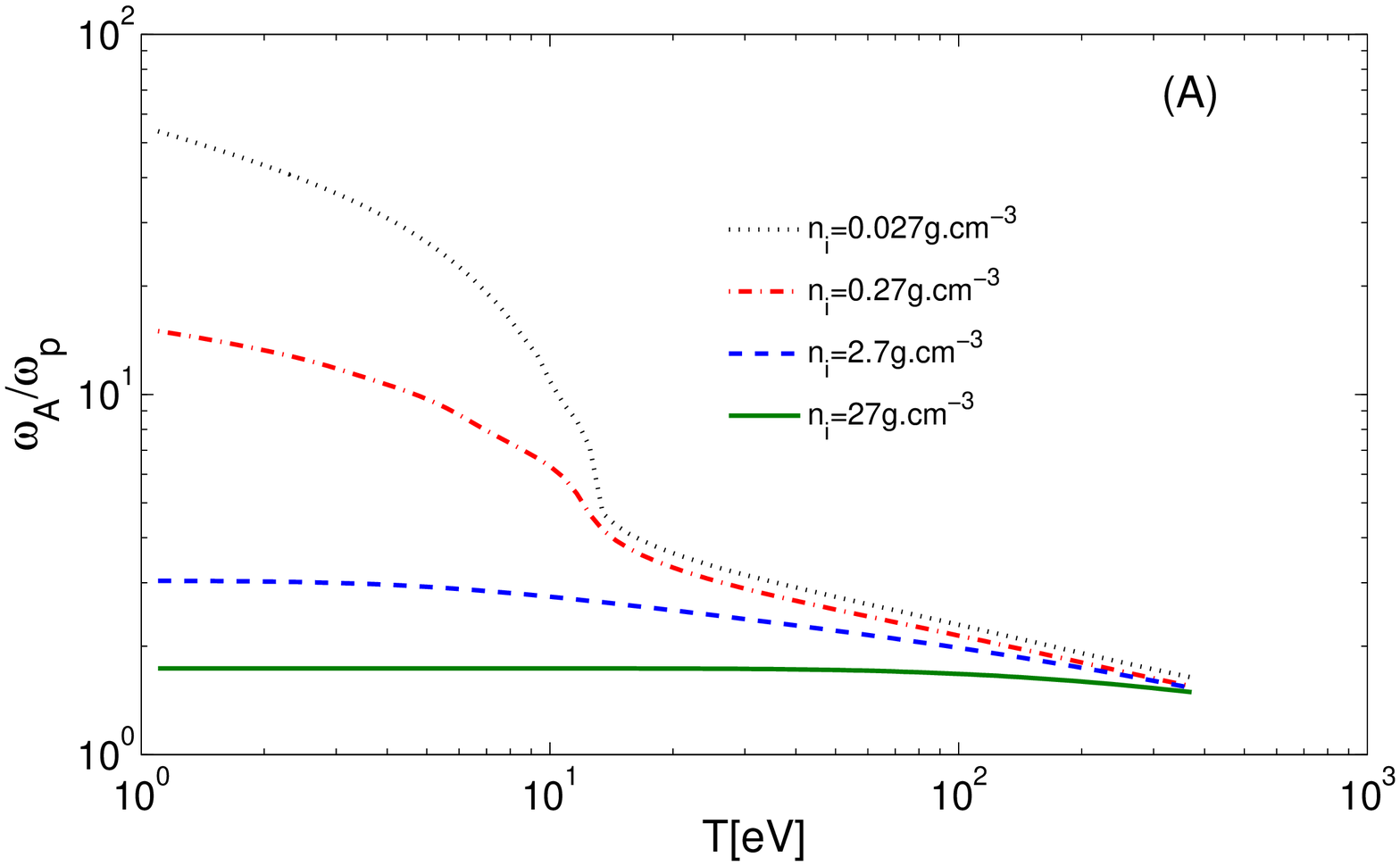} 
\includegraphics[width=8.5cm]{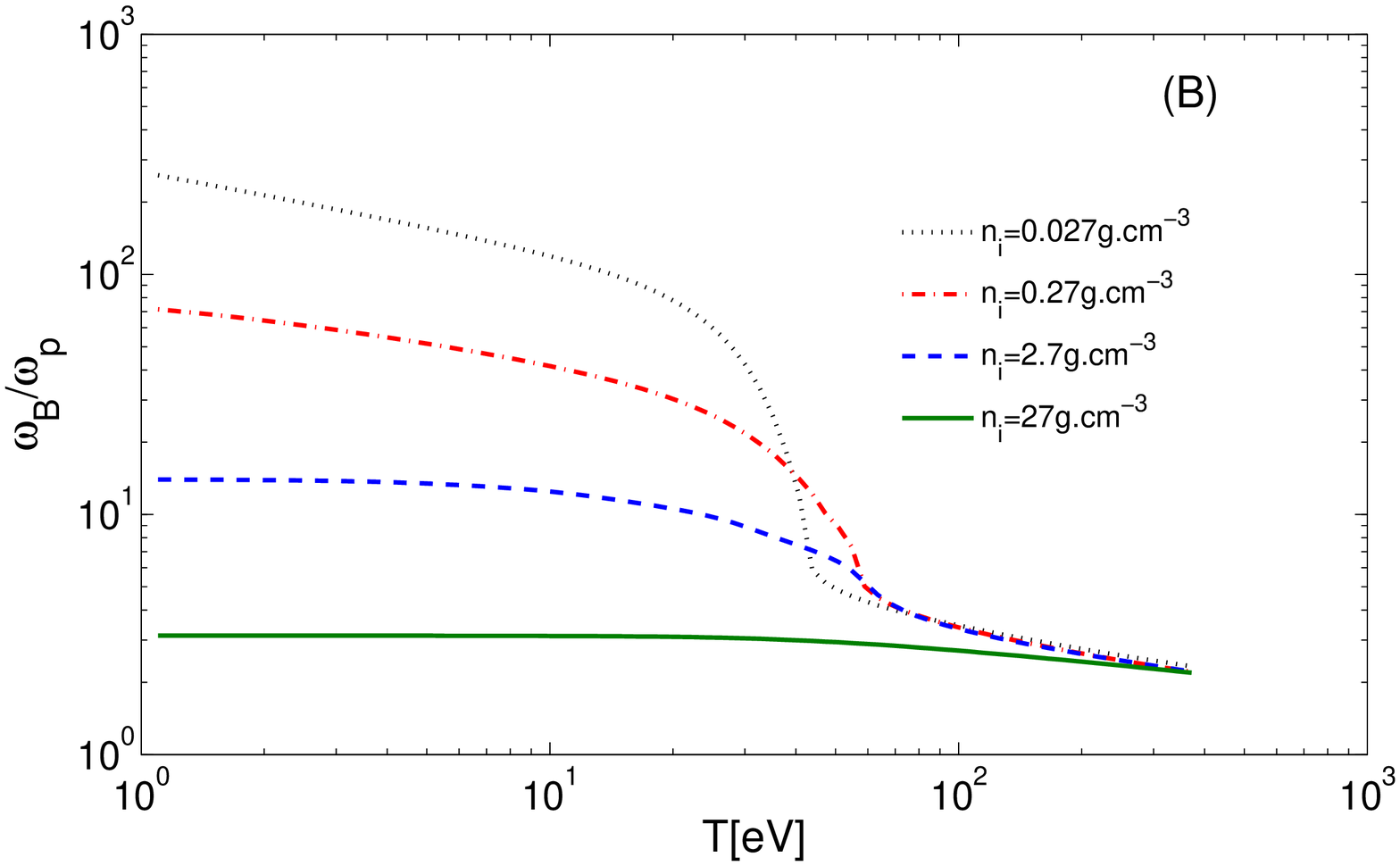}
}
\caption{(color online) (A) $\omega_{A}$, the frequency at which the \Bremsstrahlung contribution to opacity cancels, in function of temperature for four values of ion mass density. (color online) (B) $\omega_{B}$, the frequency at which the \Bremsstrahlung contribution becomes larger that the collective one, in function of temperature for four values of ion mass density.
\label{fig:AnnulationBrem}}
\end{figure}

On figures \ref{fig:AnnulationBrem}  we present the $\omega_A$ and  $\omega_B$ frequencies (in $\omega_{p}$ units) as functions of temperature for the same ion mass densities as on figure \ref{fig:ratiowp}. On these figures the plasma frequency $\omega_p$,  depends on both temperature and ion density through its dependence on the electron density $n_0^{(0)}$.

As said before the two frequencies $\omega_A$ and $\omega_B$ exist for all the physical cases studied on figure \ref{fig:AnnulationBrem}.  That means that there is always a frequency range above the plasma frequency where the collective effects dominate. In $\omega_p$ units this frequency range is particularly large at low temperatures and low ion densities. At high temperatures $\omega_A$ and $\omega_B$ decrease and tend to $\omega_p$. Thus in the high temperatures limit, the \Bremsstrahlung term tends to dominate over the whole frequency range $\omega>\omega_p$, as expected from standard theories. Hence, as already discussed the \Bremsstrahlung and the collective terms, defined in a rather formal way, behave as would be expected from standard theories. This supports our remark about the domination of the collective effects in a frequency region which is not limited to the vicinity of the plasma frequency. Therefore there is a need of self-consistent approaches to the dynamical response in plasma.

\section{Conclusion}
\label{Conclusion}

The frequency-dependent linear-response theory of the Thomas-Fermi Average-Atom (TFAA) of \cite{Feynman49} was studied using the Bloch hydrodynamics approach \cite{Bloch1933}. This model is the semiclassical version of the more general, Variational Average-Atom in Quantum Plasmas (VAAQP) approach. Energy extinction cross-sections in function of frequency were calculated from this model for aluminum plasma cases. The Thomas-Fermi Impurity in Jellium (TFIJ) model of \cite{Ishikawaall1998} was also implemented and results from these two models were compared.  Close to the plasma frequency the presence of a WS cavity which stems from the VAAQP formalism leads to values of the TFAA absorption cross-section  that are lower than those of the TFIJ model. Both models give the same unphysical $1/\omega^2$ behavior of the extinction cross section at high frequencies, which is due to the singularity of the Thomas-Fermi equilibrium density at the atom center. As shown in \cite{Ishikawaall1998b} removing the latter singularity using a  continuous central-ion charge density leads to a $1/\omega^6$ behavior of the high frequency cross section, which is also unphysical. For that reason, the Thomas-Fermi hydrodynamic approach does not seem to be realistic at high frequency and quantum extensions are highly needed.

The Ehrenfest-Type Atom-in-Plasma Sum Rule (ETAPSR), which was proposed in the quantum average-atom case \cite{Blenski06} was derived in the Thomas-Fermi-Bloch (TFB) LR. The numerical solutions allowed us to check this sum rule in both the TFAA and TFIJ models. This sum rule also allowed us to decompose the extinction cross section into two terms. In the first one appears the induced density and the gradient of the equilibrium potential. This term corresponds to the inverse \Bremsstrahlung cross-section, which would be the only term obtained in the independent electron approximation. In the second term, appears the induced potential. It can be viewed as accounting for collective effects.

In the TFB response theory, none of these terms appear to be negligible. The collective term appears to even dominate in some frequency range, not limited to the vicinity of the plasma frequency. This raises questions about the validity of the independent electron approximation in the general linear response theory. However among the difficulties of interpretation of these results is the fact that in the TFB LR, no independent particle approximation has been identified. It then appears worth to investigate the respective role of the two terms in the full quantum case, this should be the main issue of a future study. Methods and understanding of the LR theory of the TFB case presented in this work are necessary for such an extension.

\section{Acknowledgments}
One of the authors (TB) acknowledges fruitful discussions on the subjects related to this work with K. Ishikawa, B. U. Felderhof, J.-C. Pain, B. Cichocki and F. Perrot. This work 
has been partly supported by the European Communities under the contract of 
Association between EURATOM and CEA within the framework of the European 
Fusion Program.


\end{document}